\documentclass[structabstract]{aa}

\usepackage{amsmath}          
\usepackage{dcolumn}          
\usepackage{graphicx}         
\usepackage{txfonts}          
\usepackage{mathrsfs}         
\usepackage{rotating}         
\usepackage{subfigure}        
\usepackage{lscape}           
\usepackage{afterpage}        
\usepackage{xspace}           
\usepackage{natbib}           

\providecommand*{\wn}{\,cm$^{-1}$\xspace}

\newcommand{\mcl}[3]{\multicolumn{#1}{#2}{#3}}
\newcommand{\mrm}[1]{\ensuremath{\mathrm{#1}}}

\newcommand{\hoco}{\mbox{HOCO\ensuremath{^+}}\xspace}
\newcommand{\doco}{\mbox{DOCO\ensuremath{^+}}\xspace}
\newcolumntype{.}{D{.}{.}{-1}}

\begin{document}

\title{Accurate sub-millimetre rest-frequencies \\ for \hoco and \doco ions}

\author{L.~Bizzocchi\inst{1} \and V.~Lattanzi\inst{1} \and J.~Laas\inst{1} \and
        S.~Spezzano\inst{1} \and B.~M.~Giuliano\inst{1} \and D.~Prudenzano\inst{1} \and 
        C.~Endres\inst{1} \and O.~Sipil\"a\inst{1} \and P.~Caselli\inst{1}}

\institute{Center for Astrochemical Studies, 
           Max-Planck-Institut f\"ur extraterrestrische Physik,
           Gie\ss enbachstra\ss e 1, 85748 Garching (Germany)
           \email{[bizzocchi,lattanzi,jclaas,spezzano,giuliano,prudenzano,cendres,osipila,caselli]@mpe.mpg.de}
          }

\titlerunning{\hoco and \doco rest frequencies}
\authorrunning{L. Bizzocchi et al.}
%

\abstract
%
{\hoco is a polar molecule that represents a useful proxy for its parent molecule CO$_2$,
which is not directly observable in the cold interstellar medium.
This cation has been detected towards several lines of sight, including massive star forming
regions, protostars, and cold cores.
Despite the obvious astrochemical relevance, protonated CO$_2$ and its deuterated variant, 
\doco, still lack an accurate spectroscopic characterisation.
}
%
{The aim of this work is to extend the study of the ground-state pure rotational spectra 
of \hoco and \doco well into the sub-millimetre region.
}
%
{Ground-state transitions have been recorded in the laboratory using a frequency-modulation 
absorption spectrometer equipped with a free-space glow-discharge cell.
The ions were produced in a low-density, magnetically-confined plasma generated in a 
suitable gas mixture.
The ground-state spectra of \hoco and \doco have been investigated in the 213--967\,GHz 
frequency range, with the detection of 94 new rotational transitions.
Additionally, 46~line positions taken from the literature have been accurately remeasured.
}
%
{The newly-measured lines have significantly enlarged the available data sets for \hoco and 
\doco, thus enabling the determination of highly accurate rotational and centrifugal
distortion parameters.
Our analysis showed that all \hoco lines with $K_a \geq 3$ are perturbed by a ro-vibrational
interaction that couples the ground state with the $v_5=1$ vibrationally-excited state. 
This resonance has been explicitly treated in the analysis in order to obtain molecular
constants with clear physical meaning.
}
%
{The improved sets of spectroscopic parameters provide enhanced lists of very accurate,
sub-millimetre rest-frequencies of \hoco and \doco for astrophysical applications.
These new data challenges a recent tentative identification of \doco toward a pre-stellar 
core.
}


\maketitle
%

\section{Introduction} \label{sec:intro}
\indent\indent
Protonated ions were first suggested as proxies for important interstellar 
molecules by \citet{Herbst-ApJ75-NonPol}, shortly after the first detection of charged 
polyatomic species in space (HCO$^+$, \citealt{Buhl-Nat70-HCO+}; N$_2$H$^+$, 
\citealt{Turner-ApJ74-N2H+}).
These first pioneering studies demonstrated that ion--molecule reactions must occur 
in the interstellar medium (ISM), and are capable of generating ionic forms of non-polar 
molecules, such as N$_2$, C$_2$, CO$_2$, and HCCH\@.
These species are likely to be present to a large extent in the dense gas, but they 
escape radio-telescope detection owing to the lack of rotational spectra.

Carbon dioxide (CO$_2$) is widespread in space.
It is abundant in planetary atmospheres, comets, and especially interstellar ices,
where it has been extensively detected in these latter environments by ISO and 
\emph{Spitzer} telescopes towards several lines of sight 
\citep[e.g.,][]{Whitt-ApJ98-CO2,Bergin-ApJ05-CO2}.
In the solid phase, the CO$_2$:H$_2$O ratio has been observed to vary in the range of 
0.15--0.5 in molecular clouds and protostars
\citep[e.g.,][]{Boog-ARAA15-Icy,Whitt-ApJ09-CO2,Oberg-ApJ11-Ice}.
Since the abundance of CO$_2$ has been observed to be lower in the gas phase by a factor
of~100 \citep{vDish-AA96-CO2,Boonm-AA03-CO2}, its formation is thought to proceed 
on dust grains, \emph{via} UV- or cosmic-ray-induced processing of a variety of icy 
mixtures \citep{Ioppo-AA09-CR,Menne-ApJ06-CUV,Pontopp-AA03-COice,Wata-ApJ02-COice}.
Nonetheless, speculation on the possible contribution of a gas-phase formation route 
remains \citep[e.g.,][]{Sakai-ApJ08-HOCO+}, and the difficulties involved in the direct 
observation of CO$_2$ in the infrared hinder the clarification of this matter.

Protonated carbon dioxide (\hoco) provides a useful, indirect way to trace gaseous 
CO$_2$ in the ISM.
\citet{Vastel-AA16-HOCO+} constrained the CO$_2$ abundance in the L1544 pre-stellar 
core using an extensive chemical model that considered the following \hoco main 
formation channels:
\begin{subequations} \label{eq:prot-att}
 \begin{equation} \label{eq:react:HOCO+}
  \text{CO}_2 + \text{H}_3^+ \rightarrow \text{HOCO}^+ + \text{H}_2 \,, 
 \end{equation}
 \begin{equation} \label{eq:react:HCO+}
  \text{CO} + \text{H}_3^+ \rightarrow \text{HCO}^+ + \text{H}_2 \,, \quad \text{followed by}
 \end{equation}
\end{subequations}
\begin{equation} \label{eq:OHreact}
  \text{HCO}^+ + \text{OH} \rightarrow \text{HOCO}^+ + \text{H} \,.
\end{equation}
At the steady state, they derived an indirect estimate of the [CO$_2$]/[CO] 
ratio from [HOCO$^+$]/[HCO$^+$].
The assumptions involved in this approach hold for the external layers of dense cloud 
cores, where CO freeze-out rates are moderate.
The same method, in a simplified form (i.e., neglecting reaction~\eqref{eq:OHreact}), was 
also adopted by \citet{Neill-ApJ14-HEXOS} towards Srg~B2(N), and by \citet{Sakai-ApJ08-HOCO+} 
in the Class~0 protostar IRAS~04368+2557 embedded in L1527\@.

The first laboratory identification of protonated carbon dioxide was accomplished by
\citet{Bogey-AA84-HOCO+}, who observed six rotational lines of \hoco in the 350--380\,GHz 
frequency range.
This work substantiated the tentative interstellar detection proposed by 
\citet{Thadd-ApJ81-NewMol}, and an additional, independent, confirmation was provided by 
the laboratory observation of the $\nu_1$ ro-vibrational band of \hoco
\citep{Amano-JCP85-HOCO+,Amano-JCP85-HOCO+IR}.
Later, Bogey and co-workers  published two more papers about laboratory studies
in which they enlarged the frequency coverage and extended the study to the 
isotopic species \doco and HO$^{13}$CO$^+$ \citep{Bogey-JCP86-HOCO+,Bogey-JMSt88-HOCO+}.
More recently, the low-lying $J_{K_a,K_c} = 1_{0,0}-0_{0,0}$ line of \hoco was measured 
by \citet{Ohshima-CPL96-ions} using a pulsed-jet Fourier-transform microwave (FTMW) 
spectrometer.

Despite these considerable experimental efforts, the spectroscopic characterisation
of this astrophysically-relevant ion remains not fully satisfactory.
Recordings of the pure rotational spectra are indeed restricted to a rather limited 
frequency range; only a few lines have been measured in the 3\,mm band, and the whole 
spectral region above 420\,GHz is completely unexplored.
No $b$-type transitions were measured for \doco, thus resulting in a poorly 
determined $A$ rotational constant for this isotopic species.
Moreover, anomalously large centrifugal distortion effects are present in both \hoco and 
\doco.
As a result, the line positions of many astronomically important features cannot be 
computed to a desirable accuracy.

For example, the rest frequencies used by \citet{Neill-ApJ14-HEXOS} to assign the \hoco, 
$K_a = 1-0$ ladder ($J=1-7$) observed towards the Galactic centre are affected by
1$\sigma$ uncertainties of 300--400\,kHz, as indicated by the JPL line catalogue 
\citep{Pick-JQSRT98-JPL}.
Also, the tentative detection of \doco in L1544 claimed by \citet{Vastel-AA16-HOCO+} is 
based on a reference datum that is not fully reliable: the $J_{K_a,K_c} = 5_{0,5}-4_{0,4}$ 
line position provided by the JPL catalogue is $100\,359.55\pm 0.035$\,MHz, but a 
calculation performed using the ``best'' literature spectroscopic data 
\citep{Bogey-JMSt88-HOCO+} gives $100\,359.14$\,MHz.
The resulting 410\,kHz discrepancy (1.2\,km\,s$^{-1}$) therefore hints at possible issues 
affecting the spectral analysis of \doco.

With the aim of providing highly-accurate rest frequencies for astrophysical applications,
we have performed a comprehensive laboratory investigation of the pure rotational spectra
of \hoco and \doco. 
About fifty new lines were recorded for each isotopologue and, in addition, many literature 
transitions were accurately remeasured, to either refine their frequency positions or to 
rule out possible misassignments.
The measurements presented in this work also extend towards the THz region, thus 
considerably enlarging the frequency range with respect to previous studies.

The data analysis shows that the \hoco spectrum is affected by a ro-vibrational 
interaction, coupling the ground state with the low-lying $v_5=1$ vibrationally-excited
state. This resonance is characteristic of many quasi-linear molecules, such as 
HNCO \citep{Nieden-JMS95-HNCO}, HNCS \citep{Nieden-JMS95-HNCS}, and HN$_3$ 
\citep{Krakow-JMS68-qlin}.
In \hoco, this perturbation produces non-negligible effects for rotational levels having 
the quantum number $K_a$ greater than 2.
A special treatment was adopted to analyse this spectrum in order to retrieve a set of 
spectroscopic constants with clear physical meaning.

\section{Experiments} \label{sec:exp}
\indent\indent
The spectra described in this work have been recorded with the frequency-modulation (FM) 
sub-millimetre absorption spectrometer recently developed at the Center for
Astrochemical Studies (Max-Planck-Institut f\"ur extraterrestrische Physik) in Garching.

The instrument is equipped with a negative glow-discharge cell made of a Pyrex tube
(3\,m long and 5\,cm in diameter) containing two stainless steel, cylindrical hollow 
electrodes separated by 2\,m\@.
The plasma region is cooled by liquid nitrogen circulation and is contained inside a 
2\,m-long solenoid, which can produce a coaxial magnetic field up to $\sim$300\,G to 
enhance the discharge negative column \citep{DeLucia-JCP83-Glow}.

The radiation source is an active multiplier chain (Virginia Diodes) that is driven by 
a synthesizer (Keysight E8257D) operating at centimetre wavelengths.
Using a series of frequency multiplication stages, this setup provides continuous coverage
across the 82--1100\,GHz frequency range.
Accurate frequency and phase stabilisation is achieved by providing the synthesizer with 
a 10\,MHz rubidium frequency standard (Stanford Research Systems).
A liquid-He-cooled InSb hot electron bolometer (QMC Instr. Ltd.) is used as a detector.
FM is achieved by modulating the carrier signal with a sine-wave at a rate of 15\,kHz,
and then demodulating the detector output at $2f$ using a digital lock-in amplifier 
(SRS SR830). In this way, the second derivative of the actual absorption profile is 
recorded by the computer controlled acquisition system.

\hoco and \doco were produced by a DC discharge (5--15\,mA, $\sim$ 2\,kV) and a 3:1 
mixture of CO$_2$ and H$_2$/D$_2$ diluted in a buffer gas of Ar 
(total pressure $\sim$15\,$\mu$Bar).
As for other protonated ions, cell cooling is critical to enhance the absorption signals. 
In the present case, the use of a condensable precursor (CO$_2$) imposes a practical lower 
limit of $\sim 150$\,K to the cell wall temperature.
Also, magnetic plasma confinement by a $\sim$ 200\,G field was found to provide the best 
conditions for the production of the protonated CO$_2$ ion.

\begin{figure*}[!ht]
  \centering
  \includegraphics[width=1.0\textwidth]{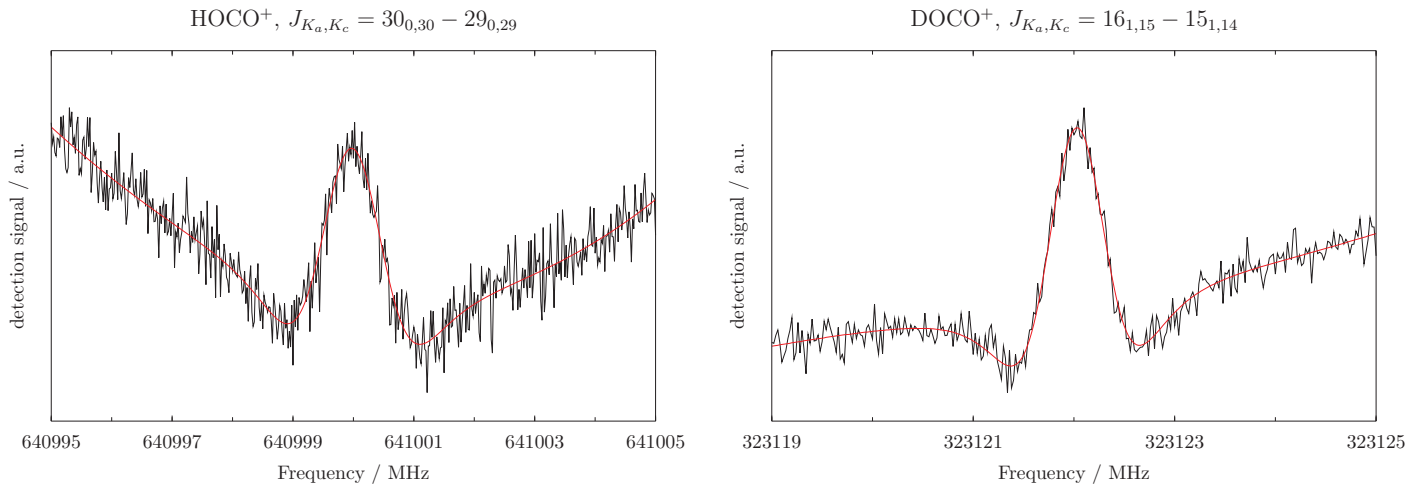}
  \caption{(\textit{left panel}) Recording of $J_{K_a,K_c} = 30_{0,30}-29_{0,29}$ transition
           of \hoco.
           Total integration time is 254\,s with 3\,ms time constant.
           (\textit{right panel}) Recording of $J_{K_a,K_c} = 16_{1,15}-15_{1,14}$ transition
           of \doco.
           Total integration time is 218\,s with 3\,ms time constant.
           (\textit{both panels}) The red traces plot the computed spectral profile obtained
           with the \texttt{proFFit} code using a modulated Voigt profile (see text).}
  \label{fig:spectra}
\end{figure*}

\section{Results and data analysis} \label{sec:res}
\indent\indent
Protonated carbon dioxide is a slightly asymmetric prolate rotor ($k = -0.9996$), with the 
$a$ inertial axis being closely aligned to the slightly bent heavy-atom backbone 
($\measuredangle(\text{O--C--O}) \approx 174^\circ$), and the hydrogen atom lying on the 
$ab$ plane ($\measuredangle(\text{H--O--C}) \approx 118^\circ$) \citep{Fort-JCP12-HOCO+}.
Hence, both $a$- and $b$-type transitions are observable. 
The electric dipole moment was theoretically computed by \citep{Green-CP76-HOCO+} yielding
$\mu_a = 2.0$\,D, and $\mu_b = 2.8$\,D.
\citet{Bogey-JMSt88-HOCO+} pointed out that these values might be inaccurate and, indeed, 
our observations of the intensity ratio between $a$- and \mbox{$b$-type} lines are not in 
agreement with the above figures.
The latest theoretical studies on \hoco \citep{Franci-JCP97-HOCO+,Fort-JCP12-HOCO+} 
do not report estimates of the dipole moments, thus we have performed an \textit{ab initio} 
calculation using the CFOUR software package\footnote{\texttt{See www.cfour.de}}. 
At the CCSD(T) level of theory \citep{Ragh-CPL89-CCSDT}, and using the cc-pCVQZ basis sets
\citep{Woon-JCP94-aug}, it yielded: $\mu_a = 2.7$\,D, and $\mu_b = 1.8$\,D.
These values give fair agreement ($\sim$ 30\%) with the line intensity ratios observed 
experimentally.

The absorption profiles of the observed transitions were modelled with the 
\mbox{\texttt{proFFit}} line profile analysis code \citep{Dore-JMS03-proFFit}, in order 
to extract their central frequency with high accuracy.
We adopted a modulated Voigt profile, and the complex component of the Fourier-transformed 
dipole correlation function (i.e. the dispersion term) was also taken into account to 
model the line asymmetry produced by the parasitic etalon effect of the absorption cell
(i.e., background standing-waves between non-perfectly transmitting windows).
The frequency accuracy is estimated to be in the range of 20--50\,kHz, depending on the
line width, the achieved signal-to-noise ratio (S/N), and the baseline.
With a magnetic field applied during the plasma discharge, the ions are produced primarily 
in the negative column, which is a nearly field-free region. 
Therefore we assume the Doppler shift caused by the drift velocity of the absorbing species 
is negligible \citep[see e.g.,][]{Tinti-ApJ07-HCO+}.

\begin{table*}[!ht]
  \caption{Assignments, measured line positions and least-squares residuals (MHz) for the
           analysed transitions of \hoco.\label{tab:lines}}
  \centering
  \begin{tabular}{ccc ccc D{.}{.}{5}D{.}{.}{-3}D{.}{.}{-3}D{.}{.}{-3} c}
  \hline\hline \noalign{\smallskip}
  $J'$ & $K'_a$ & $K'_c$ & $J$ & $K_a$ & $K_c$ &
  \multicolumn{1}{c}{observed}     &
  \multicolumn{1}{c}{o. -- c. (fit~I)}  &
  \multicolumn{1}{c}{o. -- c. (fit~II)} &
  \multicolumn{1}{c}{ass. unc.}         &
  \multicolumn{1}{c}{Ref.}         \\
  (1)  &  (2)  &  (3)  &  (4)  &  (5)  &  (6)  &  (7)  &  (8) &
  \multicolumn{1}{c}{(9)}  & \multicolumn{1}{c}{(10)}  & (11) \\
  \hline \noalign{\smallskip}
    \ldots \\
    19 & 1 & 19  &  18 & 1 & 18    &   404581.815  &  0.030  &  0.024  &  0.060  &  B86 \\
    19 & 0 & 19  &  18 & 0 & 18    &   406154.381  &  0.064  &  0.062  &  0.050  &  B86 \\
    10 & 2 &  9  &   9 & 2 &  8    &   213743.316  &  0.021  &  0.040  &  0.020  &  TW  \\
    10 & 2 &  8  &   9 & 2 &  7    &   213747.465  & -0.011  &  0.006  &  0.020  &  TW  \\
    10 & 1 &  9  &   9 & 1 &  8    &   214619.241  &  0.004  & -0.006  &  0.020  &  TW  \\
    11 & 1 & 11  &  10 & 1 & 10    &   234270.871  &  0.013  &  0.004  &  0.020  &  TW  \\
    11 & 3 &  *  &  10 & 3 &  *    &   235016.895  &         & -0.013  &  0.020  &  TW  \\
    \ldots \\[0.5ex]
  \hline
  \end{tabular}
  \tablefoot{The full table is available in electronic form at the CDS.
             Column content: (1--3) upper state rotational quantum numbers $J'_{K'a,K'c}$;
             (4--6) lower state rotational quantum numbers $J_{Ka,Kc}$;
             (7) measured line position;
             (8) least-squares residual in the fit~I (see text);
             (9) least-squares residual in the fit~II (see text);
             (10) assumed uncertainty;
             (11) Reference: TW (this work), B86 \citep{Bogey-JCP86-HOCO+}.
             Asterisks in column (3) and (6) mark unresolved asymmetry doublets.
             An empty field in column (8) indicates that the line is not considered in fit~I.
             }
\end{table*}

\subsection{\hoco} \label{sec:res:hoco+}
\indent\indent
The search for new rotational transitions of \hoco was guided by the spectroscopic 
parameters previously reported by \citet{Bogey-JMSt88-HOCO+}, thus their assignment 
was accomplished in a straightforward way.
However, at frequencies above 500\,GHz, increasingly larger discrepancies 
($\sim$ 500\,kHz) were found between observed and predicted line positions. 
Forty-three new rotational lines were recorded, reaching a maximum quantum number $J$ 
values of~29 and a frequency as high as 967\,GHz. 
These data included 12~$b$-type transitions belonging to the $^bP_{+1,-1}$, $^bQ_{+1,-1}$,
and $^bR_{+1,+1}$ branches.
In addition, 18~lines previously reported by \citet{Bogey-JCP86-HOCO+,Bogey-JMSt88-HOCO+}
were remeasured to check/improve their frequency positions. 
Figure~\ref{fig:spectra} (\textit{left panel}) shows the recording of the 
$J_{K_a,K_c} = 30_{0,30}-29_{0,29}$ line of \hoco located at ca.~640\,GHz, which is the 
highest frequency reached for $a$-type transitions.

The combined data set of literature and newly-measured lines was fitted employing 
a $S$-reduced, asymmetric rotor Hamiltonian in its $I^r$ representation 
\citep{Watson-in77-H} using Pickett's CALPGM programme suite \citep{Pick-JMS91-calpgm}.
Statistical weights ($w_i = 1/\sigma_i^2$) were adopted for each $i$-th datum to 
account for the different measurement precisions.
In our measurements, an estimated uncertainty ($\sigma_i$) of 20\,kHz is assigned 
to the $a$-type lines, whereas 50\,kHz is assigned to the weaker $b$-type transitions, 
the latter of which derive from comparatively noisier spectra. 
For the data taken from the literature, we adopted the assumed uncertainties given in 
the corresponding papers.
The complete list of the analysed rotational transitions is provided as electronic
supplementary material at the CDS\footnote{\texttt{http://cdsweb.u-strasbg.fr/}}.
An excerpt is reported in Table~\ref{tab:lines} for guidance.

The analysis clearly showed that the ground-state spectrum of \hoco is perturbed.
An anomalous slow convergence of the rotational Hamiltonian had already been noted by
\citet{Bogey-JCP86-HOCO+}, such that high-order terms were used to achieve a satisfactory 
fit of the measured frequencies.
Similar perturbations have been observed in many quasi-linear molecules, for which
HNCO (isoelectronic with \hoco) serves as a case study \citep{Nieden-JMS95-HNCO}.
These anomalies reflect the breakdown of the Watson-type asymmetric rotor Hamiltonian 
because of the accidental $\Delta K_a = \pm 1$ degeneracy occurring between ground-state 
rotational levels and those of a low-lying, totally-symmetric excited state.

We carried out the analysis of the \hoco spectrum following two different approaches. 
The first, simpler analysis was performed by applying a cut-off at $K_a = 2$. 
This excluded from the least-squares fit all the lines affected by the resonance, and 
allowed us to consider the \hoco ground state as isolated.
The $K_a = 0, 1, 2$ lines were fitted using a single state Hamiltonian, and these 
results are reported in the first column of Table~\ref{tab:fit-H} (referred to as fit~I 
hereafter).
This analysis provides a compact set of rotational parameters, including four quartic 
and two sextic centrifugal distortion constants. 
Having observed only one subset of $b$-type transitions ($K_a=0-1$), the $D_K$ constant 
could not be determined reliably, and was thus constrained to the value derived from 
previous infrared $\nu_1$ measurements \citep{Amano-JCP85-HOCO+IR}. 
The $H_J$ and $H_K$ sextic constants were also held fixed at their corresponding 
theoretically computed values \citep{Fort-JCP12-HOCO+}.

In the second stage of the analysis, the interaction coupling the ground and the 
$v_5 = 1$ states was explicitly treated, and all the available transitions ($K_a$ up to 5) 
were included in the least-squares fit. 
Assumptions for the rotational parameters of the $v_5 = 1$ state (actually unobserved) were 
derived from the ground state constants ($A_0$, $B_0$, $C_0$) and the theoretically computed 
vibration-rotation $\alpha$ constants from \citet{Fort-JCP12-HOCO+}.
An optimal fit was achieved by adjusting the same ground-state constants of the previous, 
simplified analysis, plus the resonance parameters $\eta_{12}^{ab}$ and $\eta_{12J}^{ab}$.
The quartic centrifugal distortion constant $D_K$ of the perturbing state was held fixed 
in the fit, but its value was updated iteratively until a minimum of the root mean square 
(RMS) deviation was reached.
The resulting spectroscopic constants of this analysis are gathered in the second column
of Table~\ref{tab:fit-H} (referred hereafter to as fit~II).
Full details of this analysis are given in Appendix~\ref{sec:resonance}.

\begin{table*}[tbh]
  \caption[]{Spectroscopic parameters determined for \hoco.$^{a}$}
  \label{tab:fit-H}
  \centering
  \begin{tabular}{ll ...}
    \hline \noalign{\smallskip}
    parameter &
    unit      &
    \mcl{1}{c}{fit~I}   &  
    \mcl{1}{c}{fit~II}  & 
    \mcl{1}{c}{\textit{ab initio}$^b$} \\[0.5ex]
    \hline \noalign{\smallskip}
    \textit{ground state} \\
    $A$        &  MHz  &   789\,947.786(23)    & 789\,944.610(21)    &  784\,759.5       \\
    $B$        &  MHz  &    10\,773.73262(26)  &  10\,773.68964(34)  &   10\,787.1       \\
    $C$        &  MHz  &    10\,609.43083(28)  &  10\,610.3413(23)   &   10\,623.7       \\
    $D_J$      &  kHz  &          3.49883(20)  &        3.54176(24)  &         3.433     \\
    $D_{JK}$   &  MHz  &          0.93613(12)  &        0.6454(38)   &         0.852     \\
    $D_K$      &  MHz  &     1\,123.57^c       &   1\,123.57^c       &       728.341     \\
    $d_1$      &  kHz  &         -0.05153(12)  &       -0.04555(20)  &        -0.04148   \\
    $d_2$      &  kHz  &         -0.01635(10)  &        0.01744(15)  &        -0.00914   \\
    $H_J$      &  Hz   &         -0.002^d      &       -0.002^d      &        -0.002     \\
    $H_{JK}$   &  Hz   &          0.402(83)    &        14.5(11)     &         0.885     \\
    $H_{KJ}$   &  kHz  &          3.597(28)    &        19.756(52)   &        -0.089415  \\
    $H_K$      &  MHz  &          3.418^d      &         3.418^d     &         3.418     \\[1.0ex]
    $v_5 = 1$ \\
    $A$        &  MHz  &                       & 767\,920.6^e        &  762\,735.5       \\
    $B$        &  MHz  &                       &  10\,789.9^e        &    10\,803.2      \\
    $C$        &  MHz  &                       &  10\,641.6^e        &    10\,655.0      \\ 
    $D_J$      &  kHz  &                       &        3.433^d      &          3.433    \\
    $D_{JK}$   &  MHz  &                       &        0.852^d      &          0.852    \\
    $D_K$      &  MHz  &                       &   1\,300^f          &        728.341    \\
    $d_1$      &  kHz  &                       &       -0.04148^d    &         -0.04148  \\
    $d_2$      &  kHz  &                       &       -0.00914^d    &         -0.00914  \\
    $H_J$      &  Hz   &                       &       -0.002^d      &         -0.002    \\
    $H_{JK}$   &  Hz   &                       &        0.885^d      &          0.885    \\
    $H_{KJ}$   &  kHz  &                       &       -0.089415^d   &         -0.089415 \\
    $H_K$      &  MHz  &                       &       3.418^d       &          3.418    \\[1.0ex]
    $\Delta E_v$      & \wn  &                 &     535.6^d         \\
    $\eta^{ab}_{12}$  &  MHz &                 &  3\,816.6(47)       \\
    $\eta^{abJ}_{12}$ &  MHz &                 &       0.180(11)     \\[1ex]
    no. of lines   &        &      74          &      93             \\
    $\sigma_w$     &        &       0.82       &       1.03          \\
    \hline\hline \noalign{\smallskip}
  \end{tabular}
  \begin{list}{}{}
   \item[$^a$]{See text for details on fit~I and fit~II. 
               Values in parentheses represent 1$\sigma$ uncertainties, expressed in 
               units of the last quoted digit.}
   \item[$^b$]{Theoretical values computed by \citet{Fort-JCP12-HOCO+}.}
   \item[$^c$]{Fixed at the value determined by \citet{Amano-JCP85-HOCO+IR}.}
   \item[$^d$]{Fixed at the \textit{ab initio} value.}
   \item[$^e$]{Derived from the ground state value plus \textit{ab initio} $\alpha$ \citep{Fort-JCP12-HOCO+}.}
   \item[$^f$]{Adjusted by step-by-step procedure.}
  \end{list} 
\end{table*}

\subsection{\doco} \label{sec:res:doco+}
\indent\indent
A limited portion of the millimetre spectrum of \doco had been recorded by 
\citet{Bogey-JCP86-HOCO+}, but the quality of the resulting spectral analysis was not 
completely adequate.
Indeed, in a second study, \citet{Bogey-JMSt88-HOCO+} had encountered serious difficulties 
in fitting new low-$K_a$ transitions together with the bulk of the previously measured 
data.
They had thus excluded all the $K_a=0$ lines from the final, published analysis.
This represents a severe shortage for the astronomical utility of this data, as the lines 
that may be most-detected in the ISM arise from these low energy levels. 
We have thus re-investigated carefully the rotational spectrum of \doco, focusing on 
low-$K_a$ lines that show predictions errors up to ca.~1\,MHz.

Extensive spectral searches had to be performed to identify $b$-type transitions due to 
the large error associated to the previous determination of the $A$ rotational constant.
Lines belonging to the $^bQ_{+1,-1}$ and $^bR_{+1,+1}$ branches were finally assigned 
at a distance of about 500\,MHz from the position computed from the literature data.
Ten $b$-type transitions were recorded in total.
The final \doco data set comprises 85~lines, covering the frequency range 
from~120 to~672\,GHz, with quantum number $J$ spanning values 5--30. 
The frequency list is provided as electronic supplementary material at the
CDS\footnote{\texttt{http://cdsweb.u-strasbg.fr/}}.
Figure~\ref{fig:spectra} (\textit{right panel}) shows the recording of the 
$J_{K_a,K_c} = 16_{1,15}-15_{1,14}$ line of \doco, obtained in about 3.5~min of integration
time.

Contrary to the parent isotopologue, the ground-state spectrum of \doco does not 
show evidence of perturbation.
For the deuterated variant, the $A$ rotational constant is smaller (14.4\wn compared to 
26.2\wn for \hoco), thus the $\Delta K_a =\pm 1$ quasi-degeneracy between the ground and 
$v_5 = 1$ excited state occurs at a higher value of $K_a$.
As a result, the transitions involving low energy levels are essentially unperturbed.
The analysis was therefore carried out by considering the ground vibrational state of 
\doco as isolated, and including in the fit transitions with $K_a$ up to~4.
A few, possibly perturbed lines involving $K_a = 5,6$ showed large deviations and were
then excluded from the final data set.
We adopted the same weighting scheme described in the previous subsection, where the 
assumed uncertainties ($\sigma_i$) were set to 20\,kHz and 50\,kHz for $a$- and $b$-type 
lines, respectively.
For the 3\,mm lines previously recorded by \citet{Bogey-JMSt88-HOCO+}, we retained the 
error reported by these authors.
These fit results are reported in Table~\ref{tab:fit-D}.

\begin{table}[tbh]
  \caption[]{Spectroscopic parameters determined for \doco.$^{a}$}
  \label{tab:fit-D}
  \centering
  \begin{tabular}{ll ..}
    \hline \noalign{\smallskip}
    parameter &
    unit      &
    \mcl{1}{c}{experimental}           &  
    \mcl{1}{c}{\textit{ab initio}$^b$} \\[0.5ex]
    \hline \noalign{\smallskip}
    \textit{ground state} \\
    $A$        &  MHz  &   432\,956.330(19)    &  431\,647.2       \\
    $B$        &  MHz  &    10\,163.99173(44)  &   10\,177.1       \\
    $C$        &  MHz  &    9\,908.68562(35)   &    9\,922.2       \\
    $D_J$      &  kHz  &         3.10721(20)   &         3.041     \\
    $D_{JK}$   &  MHz  &         0.326394(73)  &         0.332     \\
    $D_K$      &  MHz  &        279.05^c       &       279.05      \\
    $d_1$      &  kHz  &         -0.12542(32)  &        -0.1095    \\
    $d_2$      &  kHz  &         -0.03299(19)  &        -0.02218   \\
    $H_J$      &  Hz   &          0.0^c        &         0.0       \\
    $H_{JK}$   &  Hz   &          0.313(72)    &         0.506     \\
    $H_{KJ}$   &  Hz  &         -85.5(31)      &      -249.89      \\
    $H_K$      &  MHz  &            0.7683^c   &         0.7683    \\[1.0ex]
    no. of lines   &        &       85          \\
    $\sigma_w$     &        &       0.82        \\
    \hline\hline \noalign{\smallskip}
      \end{tabular}
      \begin{list}{}{}
   \item[$^a$]{Values in parentheses represent 1$\sigma$ uncertainties expressed in units 
               of the last quoted digit.}
   \item[$^b$]{Theoretical values computed by \citet{Fort-JCP12-HOCO+}.}
   \item[$^c$]{Held fixed.}
  \end{list} 
\end{table}

\section{Discussion} \label{sec:disc}
\indent\indent
The spectral analyses presented here, have been performed on an enlarged and improved data 
set, and yielded a more precise set of rotational and centrifugal distortion constants for 
\hoco and \doco.
In comparison to the latest literature data \citep{Bogey-JMSt88-HOCO+}, the new
spectroscopic constants presented in Tables~\ref{tab:fit-H} and~\ref{tab:fit-D} exhibit 
a significant reduction of their standard uncertainties.
The improvement is particularly relevant for \doco thanks to the 
identification of the weaker $b$-type spectrum, such that the precision of its $A$ 
rotational constant has been enhanced by a factor of 10$^4$.
As a consequence, predictive capabilities at millimetre and sub-millimetre wavelengths
have been improved.

Regarding \hoco, we have presented two different analyses. 
Fit~I, (Table~\ref{tab:fit-H}, Column 1) includes only lines originating from 
levels with $K_a \leq 2$.
Within the precision of our measurements, these transitions are essentially unaffected 
by the centrifugal distortion resonance that perturbs the ground-state spectrum
(see Appendix~\ref{sec:resonance}), and they can be treated with a standard asymmetric 
rotor Hamiltonian.
This simple solution provides reliable spectral predictions for all the transitions with 
upper-state energies $E_u/k < 220$\,K, hence it is perfectly suited to serve as a guide 
for astronomical searches of \hoco in the cold ISM.

In fit~II, all the experimental data are considered, and the interaction that couples 
the ground state with the $v_5=1$ vibrationally-excited state has been treated explicitly 
(Table~\ref{tab:fit-H}, Column 2).
Here, extensive use of the latest high-level theoretical calculations 
\citep{Fort-JCP12-HOCO+} has been made to derive reliable assumptions for those 
spectroscopic parameters which could not be directly determined from the measurements.
Though more complex, fit~II implements a more realistic representation of the 
rotational dynamics of this molecule, and yields a set of spectroscopic constants 
with clearer physical meaning.
Indeed, the anomalously large centrifugal distortion effects noted in the previous 
investigations \citep{Bogey-JCP86-HOCO+,Bogey-JMSt88-HOCO+} have been effectively 
accounted for.
No octic ($L_{JK}$) or decic ($P_{JK}$) constants were required for the analysis, 
and the agreement between experimental and \textit{ab initio} quartic centrifugal 
distortion constants is reasonably good.
On the other hand, the values determined for the $H_{JK}$ and $H_{KJ}$ sextic constants
should be considered only as effective approximations, since they include spurious 
resonance contributions not explicitly treated by this analysis.
Finally, it is to be noted that fit~I and fit~II, when limited to $K_a \leq 2$ lines, 
yield spectral predictions that are coincident within the $1\sigma$ computed 
uncertainties.

New \hoco and \doco line catalogues, based on the spectroscopic constants of
Table~\ref{tab:fit-H} (fit~I, only) and Table~\ref{tab:fit-D}, have been computed
and are provided as supplementary data available at the CDS.
These data listings include: the $1\sigma$ uncertainties (calculated taken
into account the correlations between spectroscopic constants), the upper-state energies, 
the line strength factors $S_{ij}\mu_g^2$ ($g=a,b$), and the Einstein $A$ coefficients 
for spontaneous emission,
\begin{equation} \label{eq:A}
 A_{ij} = \frac{16\pi^3 \nu^3}{3\epsilon_0 hc^3} \frac{1}{2J + 1} S_{ij}\mu_g^2 \,,
\end{equation}
where all the quantities are expressed in SI units and the line strengths $S_{ij}$ are 
obtained by projecting the squared rotation matrix onto the basis set that diagonalises 
the rotational Hamiltonian \citep{Gordy-1984}.
The computation was performed using $\mu_a = 2.7$\,D and $\mu_b = 1.8$\,D.
The compilation contains 57~lines for \hoco and 72~lines for \doco.
They are selected in the frequency range 60\,GHz $\leq\nu\leq$ 1\,THz and applying the 
cut offs $E_u/k < 100$\,K.
The precision of these rest-frequencies is at least $4\times10^{-8}$, which corresponds
to 0.011\,km\,s$^{-1}$ (or better) in unit of radial velocity.

Our analysis suggests that the line observed in L1544 by \citet{Vastel-AA16-HOCO+} and 
tentatively assigned to \doco, $J_{Ka,Kc} = 5_{0,5}-4_{0,4}$, cannot actually be attributed
to this ion.
The observed feature, once red-shifted by the 7.2\,km\,s$^{-1}$ $V_\mrm{LSR}$ of L1544, 
has a rest frequency of $100\,359.81$\,MHz, whereas our predicted line position is 
$100\,359.515\pm 0.002$\,MHz.
This 295\,kHz discrepancy corresponds to 0.88\,km\,s$^{-1}$, which is over twice
the average FWHM of the \hoco lines detected in the same source.
It is thus likely that the line tentatively detected by \citet{Vastel-AA16-HOCO+} does 
not belong to \doco and, indeed, these authors stated that this needed confirmation 
from laboratory measurements.

\section{Chemical model} \label{sec:chem}
\indent\indent
Predictions for the \hoco and \doco abundances in L1544 can be derived from the 
chemical model developed by \citet{Sipila-AA15-NH3}, which considers 
a static physical structure and an evolving gas-grain chemistry (as described in 
\citealt{Sipila-AA16-cC3H2}).
Table~\ref{tab:model} shows the model computed column densities of \hoco and \doco
for different cloud evolutionary ages averaged over a 30$^{\prime\prime}$ beam.
\citet{Vastel-AA16-HOCO+} found a \hoco column density of about 
$2\times 10^{11}$\,cm$^{-2}$\@. 
Comparison with the predicted column densities, shown in Table~\ref{tab:model} suggests 
that the best agreement is found at late evolutionary times. 
However, since above $3\times 10^5$\,yr, the model predicts too much CO freeze-out compared to 
the observations of \citet{Caselli-ApJ99-COdep}, we consider $3\times 10^5$\,yr as the best 
agreement time.
At this stage, the \doco/\hoco is about 10\% and the predicted \doco, 
$J = 5_{0,5} - 4_{0,4}$ main beam brightness temperature is 6 mK (assuming an excitation 
temperature of 8.5\,K, as deduced by \citet{Vastel-AA16-HOCO+} for their observed 
\hoco line). 

\begin{table}[tbh]
 \caption[]{Chemical model predictions for \hoco and \doco in L1544.}
 \label{tab:model}
 \centering
 \begin{tabular}{l ccc}
  \hline \noalign{\smallskip}
   time / yr                      &
   $N_\mrm{HOCO+}$ / cm$^{-2}$    &
   $N_\mrm{DOCO+}$ / cm$^{-2}$    &  
   $N_\mrm{DOCO+}/N_\mrm{HOCO+}$  \\[0.5ex]
   \hline \noalign{\smallskip}
   $5\times 10^4$ & $9.2\times 10^9$    & $3.9\times 10^8$  & 0.042  \\
   $1\times 10^5$ & $2.4\times 10^{10}$ & $2.1\times 10^9$  & 0.084  \\
   $3\times 10^5$ & $6.3\times 10^{10}$ & $6.7\times 10^9$  & 0.106  \\
   $5\times 10^5$ & $7.7\times 10^{10}$ & $6.6\times 10^9$  & 0.086  \\
   $1\times 10^6$ & $8.8\times 10^{10}$ & $5.9\times 10^9$  & 0.067  \\
   \hline\hline \noalign{\smallskip}
 \end{tabular}
\end{table}

\section{Conclusions} \label{sec:conc}
\indent\indent
This laboratory study substantially improves the spectroscopic characterisation of the
protonated CO$_2$ ion.
The spectral region sampled by the measurements has been considerably extended into the
sub-millimetre regime, reaching maximum frequencies of 967\,GHz for \hoco, and 672\,GHz
for \doco.
In addition, new recordings were obtained for most of the previously reported lines in
order to enhance their measurement precision using a sophisticated line profile analysis.
Our analysis shows that the line tentatively detected towards the pre-stellar core L1544
cannot be attributed to \doco.

\begin{acknowledgement} \label{sec:ack}
The authors wish to thank Mr. Christian Deysenroth for the thorough assistance in the 
engineering of the molecular spectroscopy laboratory at the MPE/Garching.
We are also grateful to Luca Dore for providing the \texttt{proFFit} line profile 
analysis code.
\end{acknowledgement}

\bibliographystyle{aa}
\bibliography{jsc-astro,%
              astroch,%
              molphys,%
              poly-ions,%
              asymm,%
              instrument,%
              misc}

\begin{appendix}

\section{Analysis of the \textit{ground state} $\pmb{\sim}$ $\pmb{v_5=1}$ interaction} 
\label{sec:resonance}
\indent\indent
The perturbation that affects the ground-state spectrum of \hoco is a common feature of
quasi-linear molecules, and reflects a breakdown of the Watson-type asymmetric rotor 
Hamiltonian caused by accidental rotational level degeneracies.
It was first described by \citet{Yamada-JMS80-HNCO}, and the theory has been treated in 
particular detail by \citet{Urban-JMS93-Hbreak}.

Due to the small $a$ component of the moment of inertia, $K_a + 1$ levels of the 
ground state can become close in energy with the $K_a$ levels of a low-lying, 
totally-symmetric vibrationally-excited state.
These levels can be coupled by the $\hat{H}_{12}$ term (i.e. the centrifugal distortion 
term) of the molecular Hamiltonian defined as:
\begin{equation} \label{eq:H12-gen}
 \hat{H}_{12} = -\sum_s \omega_s q_s C_s^{ab} [\hat{J}_b,\hat{J}_a]_+ \,,
\end{equation}
where the $[,]_+$ notation represents the anti-commutator, the sum runs over the 
totally-symmetric $s$ vibrational states, and $ab$ refers to the principal axis plane 
where the molecule lies.

In \hoco, the closest totally-symmetric ($A'$) state is the $v_5=1$, which is located 
ca.~536\wn above the ground state \citep{Fort-JCP12-HOCO+}. 
Therefore, given the magnitude of the $A$ rotational constant at $\approx 26$\wn, its 
$K_a$ rotational levels are crossed by the $K_a+1$ ground state ones at $K_a \approx 10$.
While the transitions are expected to show the largest deviations around this value of 
$K_a$, significant contributions are present already at $K_a\geq 3$ as sizeable
centrifugal distortion effects.

Substitution of $s=5$ to represent the $v_5$ normal mode thus reduces 
Eq.~\eqref{eq:H12-gen} to
\begin{equation} \label{eq:H12-v5}
 \hat{H}_{12} = -\omega_5 q_5 C_5^{ab} [\hat{J}_b,\hat{J}_a]_+ \,,
\end{equation}
where $\omega_5$ is the harmonic frequency, and $C_5^{ab}$ is the adimensional rotational 
derivative relative to the principal axes $a,b$ \citep{P&A-1982}.
In practice, the treatment of this resonance is accomplished by fitting the empirical 
parameters that multiply the rotational operator $J_a J_b + J_b J_a$.
This parameter has the form:
\begin{equation} \label{eq:H12-eta}
 \eta^{ab}_5 = \tfrac{1}{\sqrt{2}}\omega_5 C^{ab}_5 \,.
\end{equation}

A full analysis of this kind of ro-vibrational interaction requires, in principle, 
measurements of perturbed lines belonging to both interacting states.
For our study of \hoco, this approach is not feasible due to the lack of experimental data 
for the perturbing $v_5 = 1$ state.
Nonetheless, one can use the results of the theoretical study of \citet{Fort-JCP12-HOCO+} 
to derive reasonable assumptions for the missing data.
For this case, the rotational constants were computed from the ground state 
values of $A_0,B_0,C_0$, and the relevant \textit{ab initio} vibration-rotation interaction 
constants ($\alpha_5^A,\alpha_5^B,\alpha_5^C$). 
For the pure vibrational energy difference between the ground and $v_5 = 1$ state, we 
used the estimate of $\nu_5$ that includes quartic anharmonicity.
Finally, all five quartic and four sextic centrifugal distortion constants were assumed 
equal to the theoretically computed equilibrium values.
All these parameters were held fixed in the least-squares analysis of the ground state 
lines, in which we adjusted the same set of ground state parameters from fit~I, with the 
addition of the resonance parameter $\eta_{12}^{ab}$ and its centrifugal correction 
$\eta_{12J}^{ab}$.
By adopting this scheme, it was possible to reproduce the measured transitions for 
$K_a \leq 5$ without the need of additional high-order centrifugal distortion terms.
The fit was finally optimised by adjusting the $D_K$ constants of the $v_5=1$ states 
through a step-by-step procedure until the root-mean-square deviation was minimised.

The determination of the value of $\eta_{12}^{ab}$ (see Table~\ref{tab:fit-H}, fit~II) 
provides also an estimation of the magnitude of the adimensional resonance parameter, 
$C_5^{ab}$.
Using the theoretical $\omega_5$ value (535.6\wn, \citealt{Fort-JCP12-HOCO+}), we find
$\sim 4\times 10^{-4}$.
This value is in fair agreement to those of the isoelectronic molecules H$_2$CCO, 
H$_2$CNN ($\sim 2\times 10^{-4}$, \citealt{Urban-JMS93-Hbreak}), and HNCO 
($\sim 8\times 10^{-5}$, \citealt{Nieden-JMS95-HNCO}).

\end{appendix}

\end{document}